\documentclass[conference]{IEEEtran}
\IEEEoverridecommandlockouts
\usepackage{cite}
\usepackage{amsmath,amssymb,amsfonts,amsthm}
\usepackage{algorithmic}
\usepackage{graphicx}
\usepackage{textcomp}
\usepackage{xcolor}
\usepackage{mathtools}
\usepackage[short]{optidef}

\newtheorem{lemma}{Lemma}
\newtheorem{remark}{Remark}

\DeclareMathOperator*{\argmin}{argmin}

\def\BibTeX{{\rm B\kern-.05em{\sc i\kern-.025em b}\kern-.08em
    T\kern-.1667em\lower.7ex\hbox{E}\kern-.125emX}}
\begin{document}

\title{Optimization of Liquid Lens-based Imaging Receiver for MIMO VLC Systems\vspace{-5mm}}

\author{
\IEEEauthorblockA{\normalsize{Kapila W. S. Palitharathna}\IEEEauthorrefmark{1}, \normalsize{Christodoulos Skouroumounis}\IEEEauthorrefmark{2}, and \normalsize{Ioannis Krikidis}\IEEEauthorrefmark{1}}
\IEEEauthorblockA{\IEEEauthorrefmark{1}Department of Electrical and Computer Engineering, University of Cyprus, Nicosia, Cyprus\\
}
\IEEEauthorblockA{\IEEEauthorrefmark{2}Department of Wearables and IoT, Cyprus Research \& Innovation Center Ltd (CyRIC), Nicosia, Cyprus\\
}
\IEEEauthorblockA{Email: \{palitharathna.kapila, krikidis\}@ucy.ac.cy, c.skouroumounis@cyric.eu}\vspace{-11mm}}

\maketitle

\begin{abstract}
    In this paper, a liquid lens-based imaging receiver is proposed for multiple-input multiple-output (MIMO) visible light communication (VLC) systems. By dynamically adjusting the focal length and orientation angles of the liquid lens, the spatial correlation between MIMO channel gains is reduced, leading to enhanced bit-error rate (BER) performance. Unlike static lenses, liquid lenses offer adaptability in dynamic conditions, including user mobility and random receiver orientation. An accurate mathematical framework is developed to model the channel gains of the proposed system, and an optimization problem is formulated to minimize its BER. Due to the complexity of the resulting channel model, two lens adjustment schemes, namely, ($i$) the CLS scheme, and ($ii$) the VULO scheme are introduced. Numerical results demonstrate that the proposed liquid lens-based system offers substantial BER improvements over conventional static lens-based receivers across a wide range of random receiver orientation conditions. Specifically, at a random receiver orientation variance of $10^{\circ}$, the BER is improved from $4\times 10^{-2}$ to $5\times 10^{-4}$ by employing the proposed liquid lens.
\end{abstract}

\vspace{-5mm}
\section{Introduction}\label{sec:introduction}
\vspace{-2mm}
Next generation wireless networks are envisioned to deliver extremely high data rates, ultra-low latency, and enhanced security to support emerging services such as Industry 4.0, augmented reality, e-health, and a range of novel applications. In this context, visible light communication (VLC) is expected to play a key role in next generation wireless networks, particularly for indoor short-range communication systems. VLC offers significant advantages over conventional radio frequency (RF) communication using the extensive bandwidth available in the visible light spectrum, which spans from 380 nm to 780 nm. Specifically, in indoor point-to-point links, VLC achieves data rates on the order of terabits per second~\cite{Ghassemlooy}.

On the other hand, multiple-input multiple-output (MIMO) technology has been widely explored as an effective solution to enhance data rates in VLC systems by leveraging the inherent high directivity of light-emitting diode (LED) emission patterns along with the strategic spatial placement of LEDs and photodiodes (PDs). Among MIMO VLC systems, the spatial modulation (SM) technique has received significant attention due to its robust performance under highly correlated channel conditions~\cite{Fath_2023}. Generalized spatial modulation (GSM), a generalized form of SM, offers improved bit-error rate (BER) performance compared to conventional modulation schemes, establishing it as a promising candidate for MIMO VLC applications~\cite{chockalingam_2015}. In recent years, several studies have aimed to further improve the performance of GSM-based VLC systems~\cite{Nahhal_2021, Wang_2018, Chen_2021}. In~\cite{Nahhal_2021}, a flexible GSM scheme has been proposed, enabling dynamic adjustment of modulation orders and the number of active LEDs. A spectrally efficient hybrid dimming scheme, combining spatial-domain and time-domain dimming, has been introduced in~\cite{Wang_2018}. Furthermore,~\cite{Chen_2021} has presented a group-based LED selection technique to improve link reliability by selectively excluding under-performing LEDs from the communication process.

However, achieving full diversity and spatial multiplexing gains in conventional MIMO VLC systems remains challenging due to significant channel correlations, which are influenced by factors such as the spatial positioning of LEDs and PDs, inter-element spacing, LED radiation patterns, and the field-of-view (FoV) of the PDs~\cite{chockalingam_2015}. The inherently low inter-LED/PD spacing in practical VLC transmitters/receivers, along with the use of planar LED/PD arrays, are primary contributors to the pronounced degradation in BER performance. To mitigate correlation among the elements of the channel matrix, various techniques have been explored, including optimizing transmitter/receiver geometries, incorporating intelligent reflecting surfaces (IRSs), and employing optical lenses~\cite{Asanka_2015, Shiyuan_2023, Sushanth_2018}. In~\cite{Asanka_2015}, an angle diversity receiver utilizing pyramid/hemispherical geometries has been proposed to enhance signal reception capabilities. The integration of IRSs within MIMO VLC frameworks has been systematically analyzed in~\cite{Shiyuan_2023}. Additionally, the use of imaging receivers equipped with optical imaging lenses has been shown to effectively reduce channel correlation and improve system performance~\cite{Sushanth_2018}. 

Recent advancements in optical lens technology have introduced adaptable/tunable liquid lenses, which can significantly improve communication efficiency by dynamically optimizing optical paths and enhancing signal quality~\cite{Ngatched_2021, Zohrabi_2016, Tian_2022}. In~\cite{Ngatched_2021}, a liquid crystal-based IRS has been utilized at the receiver to dynamically steer incident light beams toward the effective area of the PD by adjusting the refractive index. Although omitted in the context of VLC systems, several innovative non-mechanical/mechanical liquid lens architectures have been proposed, capable of altering both the shape and orientation of the liquid surface~\cite{Zohrabi_2016, Tian_2022}. In~\cite{Zohrabi_2016}, the authors have demonstrated one- and two-dimensional beam steering using multiple tunable liquid lenses. In~\cite{Tian_2022}, an adaptable liquid lens has been studied which has three degrees of freedom \textit{i.e.}, focal length, azimuth angle, and polar angle. To the best of the authors' knowledge,  no prior work has investigated the use of liquid lenses to optimize the BER performance of MIMO VLC systems, particularly under dynamic conditions involving user mobility and random receiver orientations.

In this paper, we study a GSM-based indoor MIMO VLC system that incorporates a liquid convex lens-assisted imaging receiver to enhance channel gains/minimize channel correlation across a wide range of user mobility and receiver orientation conditions. A mathematical framework is established to accurately model the channel gains of the proposed system, incorporating key parameters such as user position, receiver orientation, lens focal length, and lens orientation. We formulate an optimization problem to minimize the BER of the proposed system by adjusting the focal length and the orientation angles of the liquid lens. Due to the complexity, two low-complexity optimization schemes, namely, the closest LED selection (CLS) and the vertical upward lens orientation (VULO) are proposed. Our results reveal that the use of liquid lenses in MIMO VLC systems is helpful in improving BER performance under user mobility and random receiver orientation conditions. In addition, the lens adjustment schemes presented are helpful in achieving significant performance gains under a wide range of dynamic conditions.

\begin{figure}[!t]
    \centering
    \includegraphics[width=0.95\columnwidth]{"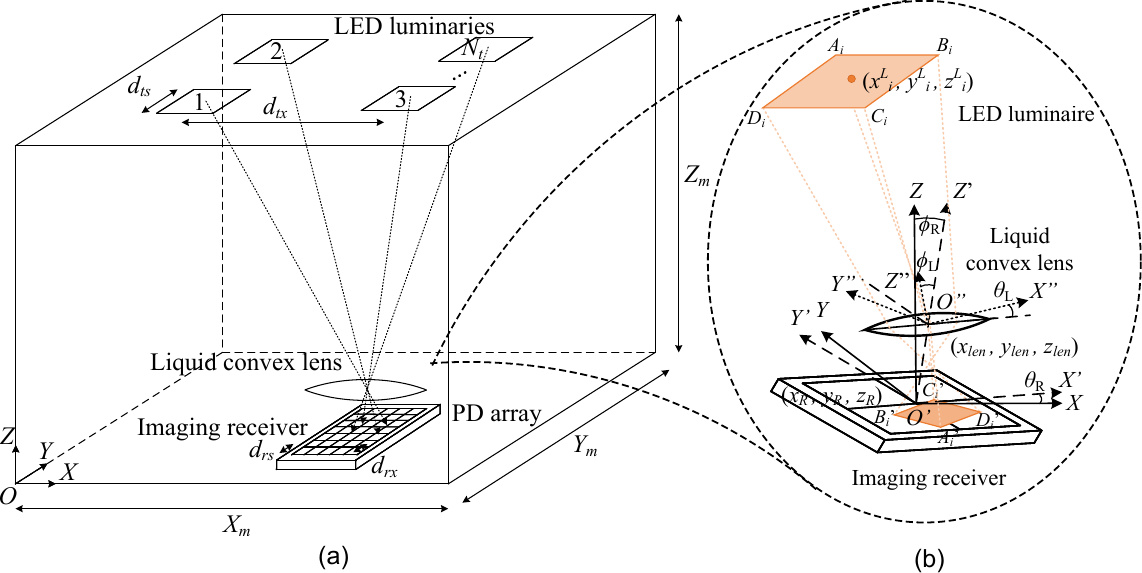"}
        \vspace{-4mm}
    \caption{(a) Liquid lens-assisted imaging receiver-based MIMO VLC system. (b) Detailed structure of the liquid lens-assisted imaging receiver.}
    \label{fig:f1}
    \vspace{-7mm}
\end{figure}

\vspace{-1mm}
\section{System Model}\label{sec:system}
\vspace{-1mm}
We consider an indoor MIMO VLC system consisting of a ceiling-mounted array of $N_t$ square LEDs, each with a side length of $d_{ts}$ and an inter-LED spacing of $d_{tx}$, which transmits data to a mobile device, as illustrated in Fig. \ref{fig:f1}(a). The global coordinate frame of the room is denoted by $OXYZ$. The geometric center of the $i$-th LED is located at $\hat{\boldsymbol{P}}_{i} \hspace{-1mm}= \hspace{-1mm}(x_i^L,y_i^L,z_i^L)$ with respect to (w.r.t.) the $OXYZ$ frame. The mobile device is equipped with an imaging receiver comprising $N_r$ square PDs, each with a side length of $d_{rs}$, arranged on a plane with an inter-PD spacing of $d_{rx}$~\cite{Sushanth_2018}. The center of the PD plane is located at $\hat{\boldsymbol{P}}_{R} \hspace{-1mm}=\hspace{-1mm}(x_R,y_R,z_R)$ w.r.t. the $OXYZ$ frame. A local coordinate frame, denoted as $O'X'Y'Z'$, is assigned to the receiver. To focus incident light spots precisely onto the PDs and to reduce inter-channel interference, a liquid lens~\cite{Sushanth_2018} is mounted on the receiver, with its centroid placed at a vertical distance of $d_{len}$ along the $O'Z'$ axis, as shown in Fig. \ref{fig:f1}(b). In this study, we consider both user mobility and random receiver orientation, where the receiver's coordinate frame $O'X'Y'Z'$ may be rotated relative to the global $OXYZ$ frame by an azimuth angle $\theta_R$ and a polar angle of $\phi_R$. 

\begin{figure}[!t]
    \centering
    \includegraphics[width=0.68\columnwidth]{"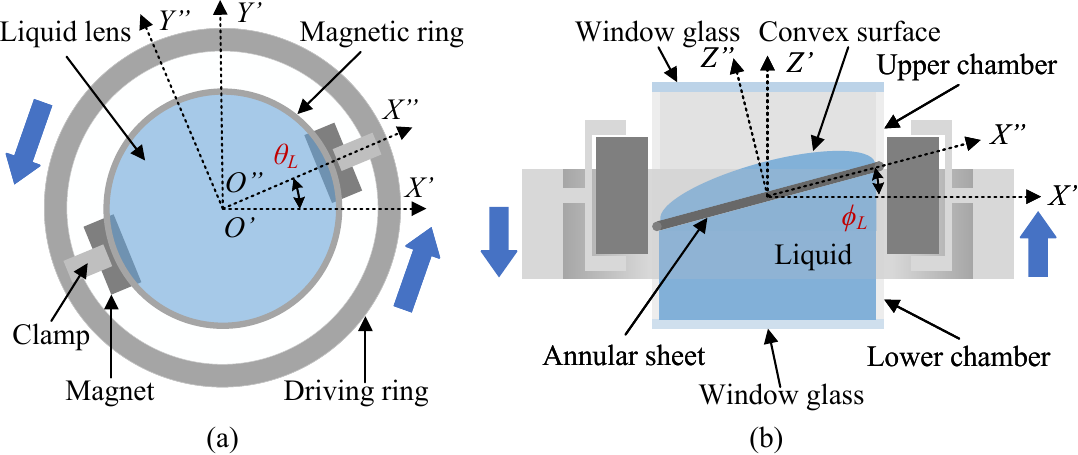"}
        \vspace{-3mm}
    \caption{Liquid lens architecture. (a). Top view. (b). Side view.}
    \label{fig:f3_lens}
        \vspace{-7mm}
\end{figure}

The liquid lens architecture considered in this work is illustrated in Fig. \ref{fig:f3_lens}. It consists of a liquid-filled chamber, where the curvature and orientation of the convex liquid surface can be dynamically controlled by adjusting the magnetic force applied to an annular metal ring on its surface~\cite{Tian_2022}. In this architecture, the focal length of the lens, denoted by $f$, is tuned by varying the vertical force exerted on the annular ring~\cite{Tian_2022}. Additionally, the azimuth angle $\theta_L$ and polar angle $\phi_L$ of the lens can be adjusted w.r.t. the receiver's local coordinate frame $O'X'Y'Z'$ by modifying the forces applied to the magnet and the driving ring. This allows precise steering and focusing of incident light spots from the LEDs onto the PDs~\cite{Tian_2022}.  Furthermore, a local coordinate frame denoted as $O''X''Y''Z''$ is established at the center of the liquid lens, which facilitates the derivation of channel gains in the proposed system. 

We use GSM to convey information in which bits are conveyed through modulation symbols sent on active LEDs as well as through LED activation patterns~\cite{chockalingam_2015}. In a single channel use, $N_a$ out of $N_t$ LEDs are activated, and each active LED emits $M$-ary intensity modulation symbol selected from the set $\mathbb{M}$. The set of intensity levels $\mathbb{M}$ includes $I_m = \frac{2I_Pm}{M+1}$, where $m = \{1,2,\ldots, M\}$, $M = |\mathbb{M}|$, and $I_P$ is the mean optical power emitted~\cite{chockalingam_2015}. Hence, the total number of bits conveyed per channel use {(bpcu)} is $\eta_{gsm} = \left\lfloor \log_2\left(^{N_t}C_{N_a}\right)\right\rfloor +N_a\lfloor\log_2 M\rfloor \text{bpcu}$~\cite{chockalingam_2015}. The transmit signal vector is $\mathbf{x} = [x_1, x_2, \ldots, x_{N_t}]^T$, where $x_i\in\{\mathbb{M}\cup0\}$ is the transmit signal at the $i$-th LED. The received signal vector $\mathbf{y}$ in the electrical domain at the receiver can be expressed as $\mathbf{y} = \alpha r\mathbf{H}\mathbf{x}+\mathbf{n}$, where $\alpha$ is the electrical-to-optical conversion efficiency of LEDs, and $r$ is the responsivity of PDs. $\mathbf{H}$ is the $N_r \times N_t$ dimentional optical channel gain matrix whose $(j,i)$-th element, $h_{i,j}$, is the optical channel gain from the $i$-th LED to the $j$-th PD, and $\mathbf{n}=[n_1, n_2, \cdots, n_{N_r}]^T$ is the noise vector, where each element $n_j$ is real AWGN noise with zero mean and variance $\sigma^2$. We use maximum likelihood (ML) detection at the receiver. Let $\mathbb{S}_{Tx} = \{\mathbf{x}_1, \mathbf{x}_2, \cdots, \mathbf{x}_L\}$ be the set of all possible transmit signal vectors for a given GSM scheme. The ML detection rule for MIMO VLC systems is
\vspace{-1mm}
\begin{equation}
    \tilde{\mathbf{x}}=\argmin_{\mathbf{x}\in \mathbb{S}_{Tx}} ||\mathbf{y}-\alpha r \mathbf{H}\mathbf{x}||^2.
    \label{equ:e3}
\end{equation}

\vspace{-0mm}
Now, we describe the light propagation model used in our system. The optical channel gain from the $i$-th LED to the $j$-th PD, denoted by $h_{i,j}$, is expressed as~\cite{Sushanth_2018}
\vspace{-2mm}
\begin{equation}
    h_{i,j} = h_{i}^{LoS}h_{i,j}^{len},
    \label{equ:e4}
    \vspace{-2mm}
\end{equation}
where $h_{i}^{LoS}$ is the line-of-sight (LoS) light propagation model from the $i$-th LED to the aperture of the imaging lens, and $h_{i,j}^{len}$ is the imaging channel gain from the $i$-th LED to the $j$-th PD. The LoS component in an indoor VLC system is deterministic and is modeled using the widely adopted Lambertian radiation model~\cite{Fath_2023}. The LoS channel gain from $i$-th LED to the centroid of the aperture of the lens can be expressed as
    \vspace{-1mm}
	\begin{equation}\label{equ:e10}
		h_{i}^{LoS}	=
		\displaystyle\frac{(m+1)A_{L}}{2\pi d_{i}^2}\cos^m(\theta_{i})\cos(\phi_{i})\Pi\left(\frac{\phi_i}{\phi_{FoV}}\right),
      \vspace{-1mm}
	\end{equation}
where $m=-\ln(2)/\ln(\cos(\theta_{1/2}))$ is the Lambertian order of the LED, $A_{L}$ is the aperture area of the lens, $d_{i}$ is the Euclidean distance between the centroid of the $i$-th LED and the centroid of the lens aperture, and $\theta_{1/2}$ is the half-power semi-angle of the LEDs. Furthermore, $\theta_{i}$ is the irradiance angle of the $i$-th LED, $\phi_{i}$ is the incident angle at the lens from the $i$-th LED, $\phi_{FoV}$ is the field-of-view (FoV) of the PD, and $\Pi(x)$ is the rectangular function. To obtain $h_{i,j}^{len}$, we use a geometric approach as explained in the Subsection \ref{subsec:h_len}.

\vspace{-1mm}
\section{MIMO VLC Channel Characterization}\label{sec:channel}
\vspace{-1mm}
We present an in-depth analysis of the channel model by employing three-dimensional geometric modeling in conjunction with geometric optics principles. Specifically, we derive the rotation matrices, the unit normal vectors of the receiver and lens, and the spatial positions of the PDs relative to the room’s coordinate frame. Finally, the LoS light propagation, $h_{i}^{LoS}$, and the imaging channel gain, $h_{i,j}^{len}$, are evaluated.
\vspace{-2mm}
\subsection{Preliminary results}
To begin with, the unit normal vector of the receiver plane, $\hat{\boldsymbol{\eta}}_{R}$, is derived in the following Lemma.
\vspace{-1mm}
\begin{lemma}\label{Lemma1}
    The unit normal vector to the PD plane, $\hat{\boldsymbol{\eta}}_{R}$, is given by
    \vspace{-3mm}
    \begin{equation}\label{equ:eta_R}
    \hat{\boldsymbol{\eta}}_{R}
    = 
    \left[\text{c}{\theta_R}\text{s}{\phi_R} \quad
    \text{s}{\theta_R}\text{s}{\phi_R} \quad 
    \text{c}{\phi_R}\right]^T,
\end{equation}
where $\theta_R$ and $\phi_R$ represent the rotation angles of the receiver's coordinate frame around its $Z$- and $Y$-axes, respectively, $\text{c}\theta = \cos{\theta}$, and $\text{s}\theta = \sin{\theta}$.
\end{lemma}
\vspace{-4mm}
\begin{proof}
    See Appendix \ref{Appendix1}.
\end{proof}
\vspace{-2mm}
In the following Remark, we compute the coordinates of the $j$-th PD \textit{i.e.}, $\hat{\boldsymbol{P}}_j$ and the center of the lens \textit{i.e.}, $\hat{\boldsymbol{P}}_{len}$, w.r.t. the room's coordinates $OXYZ$.
\begin{remark}
    The coordinates of the $j$-th PD, $\hat{\boldsymbol{P}}_j$ and the center of the lens, $\hat{\boldsymbol{P}}_{len}$, w.r.t. the frame $OXYZ$, are given by
    \begin{equation}
    \hat{\boldsymbol{P}}_j =
    \begin{bmatrix}
    x_{j}\\
    y_{j}\\
    z_{j}
    \end{bmatrix}
    = 
    \begin{bmatrix}
    x_R+x_{j}^R\text{c}{\theta_R}\text{c}{\phi_R}-y_j^{R}\text{s}{\theta_R}\\
    y_R+x_{j}^R\text{s}{\theta_R}\text{c}{\phi_R}+y_j^{R}\text{c}{\theta_R}\\ 
    z_R-x_{j}^R\text{s}{\phi_R}
    \end{bmatrix},
\end{equation}
\vspace{-3mm}
and
\vspace{-1mm}
\begin{equation}
    \hat{\boldsymbol{P}}_{len} =
    \begin{bmatrix}
    x_{len}\\
    y_{len}\\
    z_{len}
    \end{bmatrix}
    = 
    \begin{bmatrix}
    x_R+d_{len}\text{c}{\theta_R}\text{s}{\phi_R}\\
    y_R+d_{len}\text{s}{\theta_R}\text{s}{\phi_R}\\ 
    z_R+d_{len}\text{c}{\phi_R}
    \end{bmatrix},
\end{equation}
respectively, where $[x_j^R\quad y_j^R \quad 0]^T$ be the coordinates of the $j$-th PD w.r.t. receiver's coordinate frame $O'X'Y'Z'$.
\end{remark}
\vspace{-3mm}
\begin{proof}
    Let $[x_j^R\quad y_j^R \quad 0]^T$ be the coordinates of the $j$-th PD w.r.t. the frame $O'X'Y'Z'$. We assume that the $X'$ axis of the receiver lies in the direction of the user mobility. The position of the $j$-th PD w.r.t. the frame $OXYZ$ can be found by the relation $\hat{\boldsymbol{P}}_j = {^0\mathbf{R}_1^{-1}(\theta_R,\phi_R)[x_j^R\quad y_j^R \quad 0]^T+\hat{\boldsymbol{P}}_R}$. In addition, the coordinates of the center of the lens w.r.t. room's coordinate frame can be calculated as $\hat{\boldsymbol{P}}_{len} = {^0\mathbf{R}_1^{-1}(\theta_R,\phi_R)[0\quad 0 \quad d_{len}]^T+\hat{\boldsymbol{P}}_R}$.
\end{proof}
\vspace{-2mm}
The unit normal vector of the lens plane, $\hat{\boldsymbol{\eta}}_{len}$, is derived in the following Lemma.
\vspace{-3mm}
\begin{lemma}\label{Lemma2}
    The unit normal vector of the lens, $\hat{\boldsymbol{\eta}}_{len}$, is 
    \vspace{-1mm}
    \begin{equation}\label{equ:eta_len}
\hspace{-2.2mm}
    \hat{\boldsymbol{\eta}}_{len}
    = 
    \begin{bmatrix}
     \text{c}{\theta_R}\text{c}{\phi_R}\text{c}{\theta_L}\text{s}{\phi_L}-\text{s}{\theta_R}\text{s}{\theta_L}\text{s}{\phi_L}+\text{c}{\theta_R}\text{s}{\phi_R}\text{c}{\phi_L}\\
     \text{s}{\theta_R}\text{c}{\phi_R}\text{c}{\theta_L}\text{s}{\phi_L}+\text{c}{\theta_R}\text{s}{\theta_L}\text{s}{\phi_L}+\text{s}{\theta_R}\text{s}{\phi_R}\text{c}{\phi_L}\\
     -\text{s}{\phi_R}\text{c}{\theta_L}\text{s}{\phi_L}+\text{c}{\phi_R}\text{c}{\phi_L}\\
    \end{bmatrix},
    \vspace{-1mm}
\end{equation}
where $\theta_L$ and $\phi_L$ depict the rotation angles of the lens around its $Z'$- and $Y'$-axes by using its tilting mechanism.
\end{lemma}
\vspace{-3mm}
\begin{proof}
    See Appendix \ref{Appendix2}.
\end{proof}
\vspace{-3mm}
\begin{figure}[!t]
    \centering
    \includegraphics[width=0.35\columnwidth]{"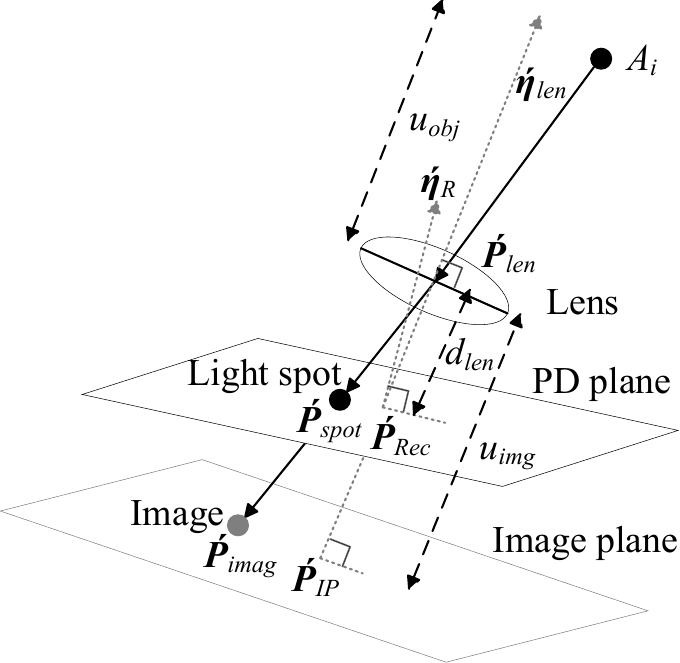"}
        \vspace{-4mm}
        \caption{Cross-sectional view of the light propagation from a point 
     on an LED and spot formation on the PD plane using the liquid lens.}
    \label{fig:f3}
        \vspace{-7mm}
\end{figure}

\vspace{-1mm}
\subsection{Calculation of $h_{i}^{LoS}$}
\vspace{-2mm}
To assess the LoS channel gain from the $i$-th LED to the centroid of the aperture of the lens, $A_L$, $d_i$, $\theta_i$, and $\phi_i$ need to be calculated. The effective aperture area of a circular lens is given by $A_L = \pi r^2$, where $r = k_{\eta}f$ is the radius of the lens as a function of $f$, and $k_{\eta}$ is a constant that depends on the refractive index of the liquid and the contact angle~\cite{Tian_2022}. The Euclidean distance between the $i$-th LED and the center of the lens is computed as $d_i= ||\hat{\boldsymbol{P}}_{i}-\hat{\boldsymbol{P}}_{len}||$. To calculate $\theta_i$ and $\phi_i$ angles, we first define the unit pointing vector from the center of the lens to the $i$-th LED as $\hat{\boldsymbol{\eta}}_{i,l}=(\hat{\boldsymbol{P}}_{i}-\hat{\boldsymbol{P}}_{len})/||\hat{\boldsymbol{P}}_{i}-\hat{\boldsymbol{P}}_{len}||$. The irradiance angle can be calculated as $\theta_i = \cos^{-1}\left(\hat{\boldsymbol{\eta}}_{i,l}\cdot[0 \quad 0 \quad 1]^T\right)$ and the incident angle is the angle between two unit vectors, $\hat{\boldsymbol{\eta}}_{i,l}$, and $\hat{\boldsymbol{\eta}}_{len}$ which is given by $\phi_i = \cos^{-1}\left(\hat{\boldsymbol{\eta}}_{i,l}\cdot\hat{\boldsymbol{\eta}}_{len}\right)$. Substituting these parameters into \eqref{equ:e10}, it can be expressed as
\vspace{-2.5mm}
	\begin{equation}\label{equ:e101}
		h_{i}^{LoS}	=
		\displaystyle\frac{k_{LoS}f^2\left(z_i^L-z_R-d_{len}c\phi_R\right)^m}{ ||\hat{\boldsymbol{P}}_{i}-\hat{\boldsymbol{P}}_{len}||^{\frac{m+3}{2}}}\left(\hat{\boldsymbol{\eta}}_{i,l}\cdot\hat{\boldsymbol{\eta}}_{len}\right),
   \vspace{-2.5mm}
	\end{equation}
where $k_{LoS} = (m+1)k_{\eta}^2/2$ is 
a constant. 
\vspace{-2.6mm}
\subsection{Calculation of $h_{i,j}^{len}$}\label{subsec:h_len}
\vspace{-2mm}
To calculate $h_{i,j}^{len}$, the projection of the $i$-th LED on the $j$-th PD is required. In the following Lemma, the coordinates of the focal point at the image plane and crossing point at the PD plane due to each vertex of $i$-th LED are derived.
 \vspace{-2mm}
\begin{lemma}\label{Lemma3}
The coordinates of the focal point at the image plane and crossing point at the PD plane due to each vertex of $i$-th LED luminary are given by
 \vspace{-3mm}
\begin{equation}\label{equ:img2}
        \hat{\boldsymbol{P}}_{PT,q_i} = \lambda_{PT}\hat{\boldsymbol{\eta}}_{ref,q_i}
 +\hat{\boldsymbol{\eta}}_{len},
  \vspace{-2mm}
    \end{equation}
where $PT\in\{img, spot\}$ denotes the image plane and the PD plane respectively, $q_i\in\{A_i,B_i,C_i,D_i\}$ are the vertices,
\vspace{-6mm}
\begin{align}
        \hat{\boldsymbol{\eta}}_{ref,q_i} &= (1/n_{l})\left[\hat{\boldsymbol{\eta}}_{len}\times (\hat{\boldsymbol{\eta}}_{len}\times \hat{\boldsymbol{\eta}}_{q_i,l}) \right] \\
        &-\hat{\boldsymbol{\eta}}_{len}\sqrt{1-(1/n_{l}^2)(\hat{\boldsymbol{\eta}}_{len}\times\hat{\boldsymbol{\eta}}_{q_i,l})\cdot(\hat{\boldsymbol{\eta}}_{len}\times\hat{\boldsymbol{\eta}}_{q_i,l})},\nonumber
        \vspace{-16mm}
    \end{align}
$n_l$ is the refractive index of the liquid,  and $\lambda_{PT}$ is a constant.
\vspace{-5mm}
\end{lemma}
\vspace{-3mm}
\begin{proof}
See Appendix \ref{Appendix3}.
\end{proof}
\vspace{-3mm}
The area of the light spot formed for the $i$-th LED on the PD plane can be expressed using the Shoelace formula\cite{Bart_1986} and is $\alpha_{i} = \left(\lvert A'_i\times B'_i\rvert +\lvert B'_i\times C'_i\rvert+ \lvert C'_i\times D'_i\rvert+\lvert D'_i\times A'_i\rvert\right)/2$. Then, $h_{i,j}^{len}=(\alpha_{i}\cap \beta_j)/\alpha_{i}$, where $\beta_j$ is the area of the $j$-th PD, and $\alpha_{i}\cap \beta_j$ is the area of intersection between the light spot generated due the $i$-th LED and the $j$-th PD.

\vspace{-2mm}
\section{BER Minimization}\label{sec:BER}
\vspace{-1mm}
We formulate an optimization problem to minimize the BER by determining the optimal focal length and orientation angles of the proposed liquid lens. The exact BER corresponding to ML detection in \eqref{equ:e3} of the GSM-based VLC system is difficult to derive in closed-form. We adopt the tight upper bound on the BER of the ML detector based on the pairwise error probability analysis as in~\cite{chockalingam_2015}, and can be expressed as
\vspace{-2mm}
    \begin{align}\label{BER}
    \hspace{-3mm}
    \resizebox{0.42\textwidth}{!}{$
        \mathrm{BER} \hspace{-0.5ex}\le \hspace{-0.5ex}\widetilde{\mathrm{BER}}
        \hspace{-0.5ex}=\hspace{-0.5ex}\displaystyle\frac{1}{\eta_{gsm}2^{\eta_{gsm}}}\hspace{-1.5ex}\sum_{m=1}^{2^{\eta_{gsm}}}\hspace{-0.5ex}
        \sum_{n=1 \atop n\ne m}^{2^{\eta_{gsm}}} \hspace{-1ex}d_\mathbf{H}(\mathbf{x}_{m},\mathbf{x}_{n}) Q\left(\hspace{-0.5ex}\frac{r\Vert \mathbf{H}(\mathbf{x}_{m}\hspace{-0.5ex}-\hspace{-0.5ex}\mathbf{x}_{n})\Vert}{2\sigma}\hspace{-0.5ex}\right)$},
        \vspace{-6mm}
    \end{align}
where $d_\mathbf{H}(\mathbf{x},\mathbf{y})$ is the Hamming distance between $\mathbf{x}$ and $\mathbf{y}$, and $Q(x)=\frac{1}{\sqrt{2\pi}}\int_{x}^{\infty}\exp\left(-\frac{u^2}{2}\right)du$ is the $Q$-function. 

We aim to minimize the BER upper bound in~\eqref{BER} w.r.t. the focal length $f$, azimuth angle $\theta_L$, and polar angle $\phi_L$ of the reconfigurable liquid convex lens. The problem can be formulated as P1 \textit{i.e.,}
    \vspace{-1mm}
	\begin{mini!}|s|
		{f,\theta_L, \phi_L}{\hspace{-3mm}\widetilde{\mathrm{BER}}(f,\theta_L,\phi_L)} 
		{\label{equ:e6}}{}
		\addConstraint{\hspace{-3mm}f\hspace{-0.5ex}\in \hspace{-0.5ex}[f^{min}\hspace{-0.5ex},f^{max}], \theta_L\hspace{-0.5ex}\in \hspace{-0.5ex}[\theta_L^{min}\hspace{-0.5ex}\hspace{-0.5ex}, \theta_L^{max}], \phi_L\hspace{-0.5ex}\in \hspace{-0.5ex}[\phi_L^{min}\hspace{-0.5ex}, \phi_L^{max}]}, \nonumber   
        \vspace{-2mm}
	\end{mini!}
where $0\hspace{-0.5ex} < \hspace{-0.5ex}f^{min}\hspace{-0.5ex}<\hspace{-0.5ex} f^{max}$ are the minimum and maximum focal lengths, $0 \hspace{-0.5ex}< \hspace{-0.5ex}\theta_L^{min} \hspace{-0.5ex}< \hspace{-0.5ex}\theta_L^{max}$ are the minimum and maximum azimuth angles, $0\hspace{-0.5ex} <\hspace{-0.5ex} \phi_L^{min}\hspace{-0.5ex}< \hspace{-0.5ex}\phi_L^{max}$ are the minimum and maximum polar angles, respectively.

Problem P1 is non-convex w.r.t. the optimization variables and can be solved using mixed monotonic programming~\cite{Matthiesen_2020}. Even though the number of optimization variables in P1 is small, this approach as well as other conventional optimization algorithms still have prohibitive computational complexity for online resource allocation. Hence, we propose two low-complexity solution approaches including ($i$) the CLS scheme, and ($ii$) the VULO scheme, and subsequently compare their performance with exhaustive search approach.

\subsection{CLS Scheme}
We present the CLS scheme, motivated by the fact that the objective function of problem P1 is equivalent to maximizing the term, $\Vert \mathbf{H}(\mathbf{x}_{m}-\mathbf{x}_{n})\Vert$. Hence, the sparsity of $\mathbf{H}(\alpha,\theta_L,\phi_L)$ needs to be increased. Furthermore, we observe that the objective function is often decreased the most by increasing the largest value of $\mathbf{H}(\alpha,\theta_L,\phi_L)$ even further. Based on this observation, we adjust the orientation angles of the lens, $\theta_L$, and $\phi_L$, such that the axis of the lens crosses the closest LED to the lens that satisfies $i^*=\argmin_{i}\ d_{i}$.

In the following Lemma, analytical expressions for the orientation angles are derived, conditioned on the location of the closest LED.
\vspace{-2mm}
\begin{lemma}\label{Lemma4}
In the proposed CLS scheme, the rotation angle of the lens around the $Y'$-and $Z'$-axes are given by
\vspace{-1mm}
\begin{align}\label{eqs:quad1_sol_sim}
\resizebox{0.23\textwidth}{!}{$
\phi_L = \cos^{-1}{\left(\text{c}{\phi_R}\left(\frac{z_{i^*}-z_{len}}{d_{i^*}}\right)\right)}$},
\end{align}
\vspace{-4mm}
and
\vspace{-3mm}
\begin{align}\label{eqs:quad1_sol_sim2}
    \resizebox{0.33\textwidth}{!}{$
    \displaystyle \theta_L = \sin^{-1}\left(\frac{\text{c}{\theta_R} \left(\frac{y_{i^*}-y_{len}}{d_{i^*}}\right)- \text{s}{\theta_R}\left(\frac{x_{i^*}-x_{len}}{d_{i^*}}\right)}{\sqrt{1-\text{c}{\phi_R}\left(\frac{z_{i^*}-z_{len}}{d_{i^*}}\right)^2}}\right)$},
\end{align}
\vspace{-1mm}
\hspace{-5mm}respectively, where $(x_{i^*}, y_{i^*},z_{i^*})$ depict the coordinate of the selected $i^*$-th LED, and $d_{i^*}= \sqrt{(x_{i^*}-x_{len})^2+(y_{i^*}-y_{len})^2+(z_{i^*}-z_{len})^2}$ denotes the distance to the closest LED from the lens.
\end{lemma}
\vspace{-2mm}
\begin{proof}
    See Appendix \ref{Appendix4}.
\end{proof}
\vspace{-2mm}

In the context of the proposed CLS scheme, $f$ is adjusted such that the distance between the lens and the light spot formed due to the closest LED is equal to $f$. This selection is motivated, since the channel gain from the closest LED to a PD is increased and, as a result, the norm in the BER expression can be increased to improve the BER value. Hence, the selected focal length in this scheme is $f^* = d_{len}/(\hat{\boldsymbol{\eta}}_{len}\cdot \hat{\boldsymbol{\eta}}_{R})$. With the help of \eqref{equ:eta_R}, and \eqref{equ:eta_len}, $f^*$ can be re-expressed as
\vspace{-1mm}
\begin{align}\label{eqs:focal_len2}
\hspace{-3.5mm}
\resizebox{0.45\textwidth}{!}{$
f^* \hspace{-0.5ex}= \hspace{-0.5ex}\frac{d_{len}}{\text{c}{\theta_R}\text{s}{\phi_R}\left(\hspace{-0.5ex}\frac{x_{i^*}-x_{len}}{d_{i^*}}\hspace{-0.5ex}\right)+ \text{s}{\theta_R}\text{s}{\phi_R}\left(\hspace{-0.5ex}\frac{y_{i^*}-y_{len}}{d_{i^*}}\hspace{-0.5ex}\right)+\text{c}{\phi_R}\left(\hspace{-0.5ex}\frac{z_{i^*}-z_{len}}{d_{i^*}}\hspace{-0.5ex}\right)}$}.
\end{align}

\vspace{-2mm}
\subsection{VULO Scheme}
In this scheme, we consider orienting the axis of the lens of the receiver vertically upward along the $z$ axis of the room coordinate frame, which can be expressed as $\hat{\boldsymbol{\eta}}_{len} = [0 \quad 0 \quad 1]^T$. By obtaining three expressions for $x$, $y$, $z$ direction vectors, the following equations hold \textit{i.e.},
\vspace{-2mm}
\begin{align}\label{eqs:362}
\text{c} {\theta_R}\text{c}{\phi_R}\text{c}{\theta_L}\text{s}{\phi_L}-\text{s}{\theta_R}\text{s}{\theta_L}\text{s}{\phi_L}+
\text{c}{\theta_R}\text{s}{\phi_R}\text{c}{\phi_L} = 0. 
\end{align}
\vspace{-8mm}
\begin{align}\label{eqs:372}
\text{s}{\theta_R}\text{c}{\phi_R}\text{c}{\theta_L}\text{s}{\phi_L}+\text{c}{\theta_R}\text{s}{\theta_L}\text{s}{\phi_L}+
\text{s}{\theta_R}\text{s}{\phi_R}\text{c}{\phi_L} = 0.
\end{align}
\vspace{-8mm}
\begin{align}\label{eqs:382}
-\text{s}{\phi_R}\text{c}{\theta_L}\text{s}{\phi_L}+\text{c}{\phi_R}\text{c}{\phi_L} = 1.
\vspace{-2mm}
\end{align}
Next, we multiply \eqref{eqs:362} by $\text{c}\theta_R$ and \eqref{eqs:372} by $\text{s}\theta_R$, and then, by taking the addition of two resulting expressions to obtain the expression $\text{c}{\phi_R} \text{c}{\theta_L} \text{s}{\phi_L}+ \text{s}{\phi_R} \text{c}{\theta_L} = 0$.
By substituting \eqref{eqs:382} into this result, we get two solutions such as \textit{Case 1:} $\phi_L = \phi_R$, and \textit{Case 2:} $\phi_L = -\phi_R$. Next, by substituting $\phi_R$ to \eqref{eqs:382}, the corresponding $\theta_L$ values can be obtained as \textit{Case 1:} $\theta_L = 180^{\circ}$, and \textit{Case 2:} $\theta_L = 0$.

In this scheme, $f^*$ is set equal to the distance from the lens to the receiver plane along the $z$ axis of the $OXYZ$ frame. In this scheme, the vertical downward light coming to the lens is focused well on the receiver plane in this scheme. The selected focus length in this scheme can be expressed as $f^* = \frac{d_{len}}{\cos{\phi_R}}$.

\vspace{-1mm}
\section{Numerical Results and Discussion}\label{sec:results}
 \vspace{-1mm}
We present numerical results to verify the performance gains of the liquid lens-assisted imaging receiver-based MIMO VLC system. A room dimension of $5\text{m}\times 5\text{m}\times3.5\text{m}$ is assumed. Unless stated explicitly, otherwise in all simulations, we have set the parameter values to, $\rho=1.5$, $A_{L}=10^{-4}$ m$^{2}$, $\Phi_{FoV} = 90^\circ$, $\theta_{1/2}=60^\circ$, $r=0.75$~A/W, $N_t = N_r=16$, $\alpha=1$ W/A, $d_{len}=0.02$~m, $d_s=0.25$~m, $d_{tx}=0.5$~m, $d_{rx}=5$ mm, $M=2$, $N_a=2$, $\sigma=10^{-6}$, $I_P=1$, $f^{min}=1$ cm, $f^{max}=15$ cm, $\theta_L^{min}=0^\circ$, $\theta_L^{max}=360^\circ$, $\phi_{min}=0^\circ$, $\phi_{max}=30^\circ$, and $k_{\eta}=0.1$.

\begin{figure}[!t]
    \centering
    \includegraphics[width=0.68\columnwidth]{"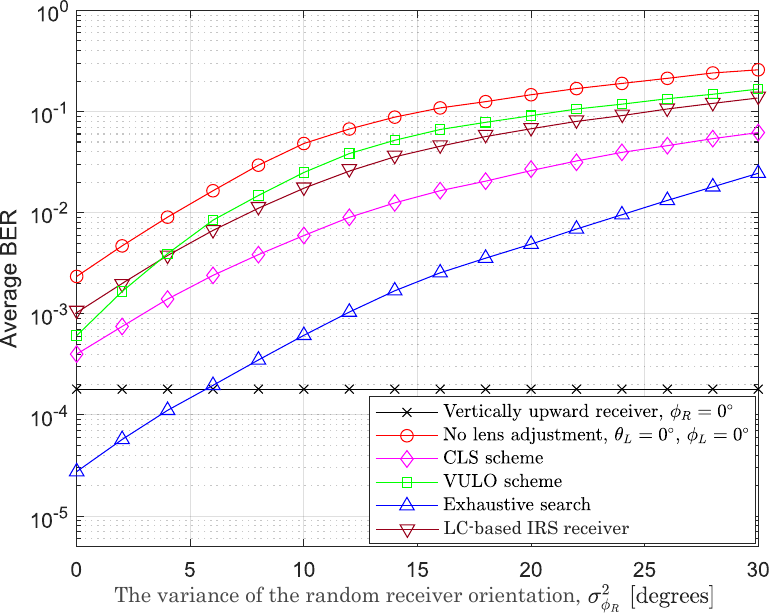"}
     \vspace{-3mm}
    \caption{BER vs. $\sigma_{\phi_R}^2$ for different optimization schemes.}
    \label{fig:result3}
    \vspace{-5mm}
\end{figure}

Fig. \ref{fig:result3} illustrates the BER performance versus the variance of the random receiver orientation, $\sigma_{\phi_R}^2$, for proposed optimization schemes, and benchmark schemes. A high $\sigma_{\phi_R}^2$ value results in a weak BER performance in all optimization schemes as the channel gain values are small, and hence, the norm in the BER expression becomes small. With no lens adjustment, the BER performance is weak compared to all other schemes and saturates fast around $\sigma_{\phi_R}^2 = 15^{\circ}$ as light spots are not focused properly on each PD. The results are helpful to identify that the use of liquid lens and lens parameters optimization results in better performance in a wide range of $\sigma_{\phi_R}^2$. In particular, a BER improvement of more than $6.7$ dB can be observed for $\sigma_{\phi_R}^2 \le 30^{\circ}$ using the CLS scheme. In addition, for $\sigma_{\phi_R}^2 
\le 6^{\circ}$, even a better performance than a static receiver is observed with liquid lens and under receiver orientation using exhaustive search. In particular, CLS and VULO schemes have similar performance gains for low $\sigma_{\phi_R}^2$ values as both schemes results in approximately equal lens parameters. Our schemes can achieve better BER than liquid crystal-based IRS receiver presented in \cite{Ngatched_2021} except VULO scheme and no lens adjustment scheme, due to lens's ability to mitigate correlation of channel gains efficiently.

\begin{figure}[!t]
    \centering
    \includegraphics[width=0.68\columnwidth]{"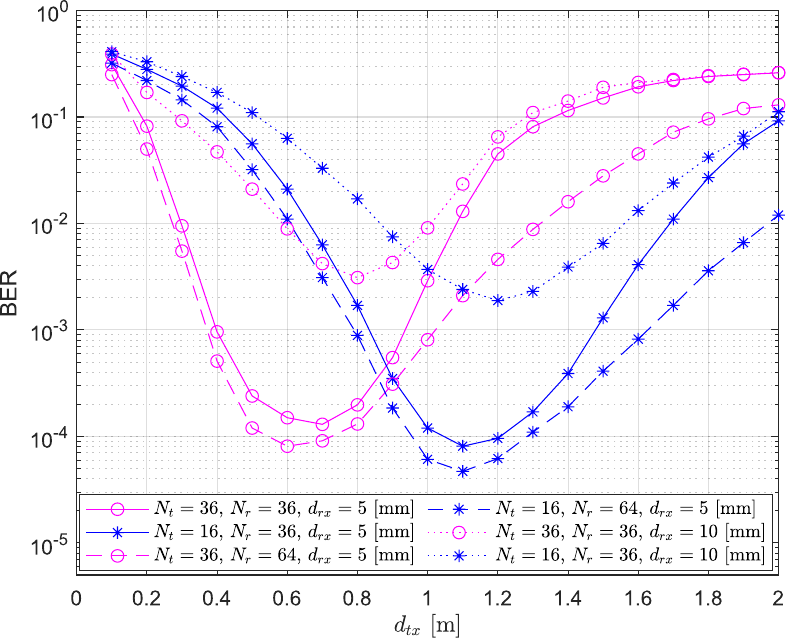"}
     \vspace{-3mm}
    \caption{BER vs. $d_{tx}$. $\sigma^2_{\phi_R} = 10^{\circ}$, and $N_a = 1$.}
    \label{fig:result4}
    \vspace{-7mm}
\end{figure}

Fig. \ref{fig:result4} shows the BER performance versus inter-LED distance, $d_{tx}$. For this simulation, we used $\sigma^2_{\phi_R} = 10^{\circ}$, $N_a = 1$. Results are illustrated for different number of LED transmitters, $N_t$, different number of PD receivers, $N_r$, and different inter-PD distance, $d_{rx}$ values. Our results show that for a certain $N_t$, there is an optimal $d_{tx}$ value that minimizes the BER. Specifically, optimal $d_{tx}$ is $0.7$ m and $1.1$ m in the cases with $N_t = 36$ and $N_t = 16$. The optimal $d_{tx}$ tends to decrease as $N_t$ increases. At low $d_{tx}$, poor BER performance is observed since spot locations on the PD plane overlap each other. Similarly, high $d_{tx}$ also results in poor BER, especially at high $N_t$, as channel gains from edge LEDs become extremely poor. Moreover, it is noted that a higher $N_r$ value results in better BER performance, especially at high $d_{rx}$ due to the fact that more PDs can capture light coming from LEDs at the corner of the room. In addition, high $d_{rx}$ results in poor performance as a sparsely placed PD array cannot capture formed light spots from the liquid lens efficiently. 

\vspace{-2mm}
\section{Conclusion}\label{sec:conclusion}
\vspace{-2mm}
In this paper, we proposed an adjustable liquid lens-assisted imaging receiver for the GSM-based MIMO VLC system. Specifically, dynamic conditions such as user mobility and random receiver orientation were taken into account. A channel gain model applicable to the proposed system is presented. We formulated an optimization problem to minimize the BER by adjusting the focal length and orientation angles of the liquid lens. Due to the mathematical complexity of the channel model, two lens adjustment schemes, including ($i$) the CLS scheme, and ($ii$) the VULO scheme were presented and compared with benchmark schemes. Our results show that the proposed liquid lens-based imaging receiver and the proposed optimization schemes are helpful in achieving superior BER performance. Specifically, at a random receiver orientation variance of $10^{\circ}$, the BER is improved from $4\times 10^{-2}$ to $5\times 10^{-4}$ by employing the proposed liquid lens. 

\vspace{-2mm}
\appendices
\section{Proof of Lemma \ref{Lemma1}}\label{Appendix1}
\vspace{-1mm}
The receiver's coordinate frame, $O'X'Y'Z'$ can be considered as a rotated version of the room's coordinate frame, $OXYZ$ by $\theta_R$ around its $Z$-axis and $\phi_R$ by its $Y$-axis. The rotation matrix between the frames $OXYZ$ and $O'X'Y'Z'$ can be expressed as  $^0\mathbf{R}_1(\theta_R,\phi_R)=\mathbf{R}_Y(\phi_R)\mathbf{R}_Z(\theta_R)$, where $\mathbf{R}_Y(\phi_R)$ is the rotation matrix around $Y$-axis by $\phi_R$ angle and $\mathbf{R}_Z(\theta_R)$ is the rotation matrix around $Z$-axis by $\theta_R$ angle and it reduces to
\vspace{-3mm}
\begin{equation}\label{equ:rot01}
\vspace{-2mm}
    ^0\mathbf{R}_1(\theta_R,\phi_R)
    = 
    \begin{bmatrix}
     \text{c}{\theta_R}\text{c}{\phi_R} & \text{s}{\theta_R}\text{c}{\phi_R} & -\text{s}{\phi_R}\\
     -\text{s}{\theta_R} & \text{c}{\theta_R}& 0\\
    \text{c}{\theta_R}\text{s}(\phi_R) & \text{s}{\theta_R}\text{s}{\phi_R} & \text{c}{\phi_R}\\
    \end{bmatrix},
\end{equation}
where $\text{c}\theta = \cos{\theta}$, and $\text{s}\theta = \sin{\theta}$. The unit normal vector to the PD plane lies along the axis, $Z'$, and hence, by using the passive rotation between two coordinate frames $OXYZ$ and $O'X'Y'Z'$, the final expression can be derived by evaluating the expression $\hat{\boldsymbol{\eta}}_{R}={^0\mathbf{R}_1^{-1}(\theta_R,\phi_R)[0\quad 0 \quad 1]^T}$.

\vspace{-2mm}
\section{Proof of Lemma \ref{Lemma2}}\label{Appendix2}
\vspace{-1mm}
The lens has two degrees of freedom, having the ability to rotate around its $Z'$-axis to change its yaw angle $\theta_L$ by rotating its magnetic ring. It can be tilted around its $Y'$-axis to create a polar angle $\phi_L$ using its tilting mechanism. Taking these into account, the rotation matrix from the receiver coordinate frame to the lens coordinate frame can be expressed as $^1\mathbf{R}_2(\theta_R,\phi_R)=\mathbf{R}_{Y'}(\phi_L)\mathbf{R}_{Z'}(\theta_L)$, where $\mathbf{R}_{Y'}(\phi_L)$ is the rotation matrix around $Y'$-axis by $\phi_L$ angle and $\mathbf{R}_{Z'}(\theta_L)$ is the rotation matrix around $Z'$-axis by $\theta_L$ angle. The unit normal vector of the lens can be thought of as a unit vector along its own $Z''$-axis, and it can be represented in the room's coordinate frame as $\hat{\boldsymbol{\eta}}_{len}=^0\mathbf{R}_1^{-1}(\theta_R,\phi_R)^1\mathbf{R}_2^{-1}(\theta_L,\phi_L)[0\quad 0 \quad 1]^T$.

\vspace{-2mm}
\section{Proof of Lemma \ref{Lemma3}}\label{Appendix3}
\vspace{-1mm}
Let $A_i$, $B_i$, $C_i$, and $D_i$ represent the vertices of the $i$-th LED transmitter, with corresponding image points on the PD plane as $A'_i$, $B'_i$, $C'_i$, and $D'_i$ (See Fig. \ref{fig:f1}(b)). The points $A_i$, $B_i$, $C_i$, and $D_i$ can be expressed as $\boldsymbol{P}_{A_i} = (x_i^L-d_{ts}/2, y_i^L-d_{ts}/2, z_i^L)$, $\boldsymbol{P}_{B_i} = (x_i^L-d_{ts}/2, y_i^L+d_{ts}/2, z_i^L)$, $\boldsymbol{P}_{C_i} = (x_i^L+d_{ts}/2, y_i^L+d_{ts}/2, z_i^L)$, and $\boldsymbol{P}_{D_i} = (x_i^L+d_{ts}/2, y_i^L-d_{ts}/2, z_i^L)$. The unit vector along a direction from the midpoint of the lens to the point $q_i\in\{A_i,B_i,C_i,D_i\}$ is given by
$\hat{\boldsymbol{\eta}}_{q_i,l}=\frac{\hat{\boldsymbol{P}}_{q_i}-\hat{\boldsymbol{P}}_{len}}{||\hat{\boldsymbol{P}}_{q_i}-\hat{\boldsymbol{P}}_{len}||}$.
Similar to the case of $\phi_i$, the incident angle on the lens from the $q_i$-th point can be expressed as $\phi_{q_i} = \cos^{-1}\left(\hat{\boldsymbol{\eta}}_{q_i,l}\cdot\hat{\boldsymbol{\eta}}_{len}\right)$. Moreover, the projection of the line connecting the mid-point of the lens to the point $q_i$ along the axis $\hat{\boldsymbol{\eta}}_{len}$ is $d_{q_i, \hat{\boldsymbol{\eta}}_{len}} = ||(\hat{\boldsymbol{P}}_{q_i}-\hat{\boldsymbol{P}}_{len})|| \cos\left(\phi_{q_i}\right)$.

Assuming the use of the paraxial optical lens, the image formed by the lens can be geometrically distortionless in comparison to the source. Let's assume the paraxial distance to the object from the lens is $u_{obj} = d_{q_i, \hat{\boldsymbol{\eta}}_{len}}$, and the paraxial distance to the image plane is $u_{img}$. By using the basic lens maker's law given by $\frac{1}{f} = \frac{1}{u_{img}}-\frac{1}{u_{obj}}$, it can be rearranged as $\frac{u_{obj}}{u_{img}} = \frac{u_{obj}}{f}+1$. This expression can be used to calculate the the magnification produced by the lens which can be expressed as $m = \frac{u_{img}}{u_{obj}} = \frac{f}{f+d_{q_i, \hat{\boldsymbol{\eta}}_{len}}}$. The cross point of the axis of the lens with the image plane is expressed as
\vspace{-2mm}
    \begin{equation}\label{equ:pip}
        \hat{\boldsymbol{P}}_{IP} = \hat{\boldsymbol{P}}_{len} - m \left(\hat{\boldsymbol{P}}_{T,q_i}-\hat{\boldsymbol{P}}_{len}\right),
        \vspace{-3mm}
    \end{equation}
where $\hat{\boldsymbol{P}}_{T,q_i} = \hat{\boldsymbol{P}}_{len}+d_{q_i, \hat{\boldsymbol{\eta}}_{len}}\hat{\boldsymbol{\eta}}_{len}$ is the position vector of the point on the axis of the lens project by the point $q_i$. The image plane is perpendicular to the axis of the lens, and hence, the position of the focal point on the image plane is 
\vspace{-2mm}
    \begin{equation}\label{equ:img}
        \left(\hat{\boldsymbol{P}}_{img,q_i} - \hat{\boldsymbol{P}}_{IP}\right)\cdot\hat{\boldsymbol{\eta}}_{len} = 0.
                \vspace{-3mm}
    \end{equation}
Considering the small area of the lens and approximating to a single reflection from the lens surface, the direction of the refracted light beam can be expressed as~\cite{Amit_2012}
\vspace{-2mm}
    \begin{align}
        \hat{\boldsymbol{\eta}}_{ref,q_i} &= (1/n_{l})\left[\hat{\boldsymbol{\eta}}_{len}\times (\hat{\boldsymbol{\eta}}_{len}\times \hat{\boldsymbol{\eta}}_{q_i,l}) \right] \\
        &-\hat{\boldsymbol{\eta}}_{len}\sqrt{1-(1/n_{l}^2)(\hat{\boldsymbol{\eta}}_{len}\times\hat{\boldsymbol{\eta}}_{q_i,l})\cdot(\hat{\boldsymbol{\eta}}_{len}\times\hat{\boldsymbol{\eta}}_{q_i,l})},\nonumber
        \vspace{-13mm}
    \end{align}
where $n_l$ is the relative refractive index of the liquid. We assume that the PD plane is close to the image plane and a single crossing point of the light with the PD plane is considered. Hence, $\hat{\boldsymbol{P}}_{img,q_i}$ can be further expressed as    
\vspace{-1mm}
    \begin{equation}\label{equ:img2}
        \hat{\boldsymbol{P}}_{img,q_i} = \lambda_{ref}\hat{\boldsymbol{\eta}}_{ref,q_i}
 +\hat{\boldsymbol{\eta}}_{len},
 \vspace{-2mm}
    \end{equation}
where $\lambda_{ref}$ is a constant. Substituting \eqref{equ:img2} and \eqref{equ:pip} into \eqref{equ:img}, $\lambda_{ref}$ can be obtained, and hence, $\hat{\boldsymbol{P}}_{img,q_i}$ can be resolved. The locations of the light spots on the PD plane can be obtained as follows. 
The PD plane is perpendicular to the axis of the receiver, and hence, the position of the light spot on the PD plane is expressed as 
\vspace{-2mm}
    \begin{equation}\label{equ:pdplane}
        \left(\hat{\boldsymbol{P}}_{spot,q_i} - \hat{\boldsymbol{P}}_{R}\right)\cdot\hat{\boldsymbol{\eta}}_{Rec} = 0.
        \vspace{-2mm}
    \end{equation}
Also, $\hat{\boldsymbol{P}}_{spot,q_i}$ lies on the reflected ray and is expressed as
\vspace{-2mm}
    \begin{equation}\label{equ:img3}
        \hat{\boldsymbol{P}}_{spot,q_i} = \lambda_{spot}\hat{\boldsymbol{\eta}}_{ref,q_i}
 +\hat{\boldsymbol{\eta}}_{len},
 \vspace{-2mm}
    \end{equation}
where $\lambda_{spot}$ is a constant. Substituting \eqref{equ:img3} into \eqref{equ:img}, $\lambda_{spot}$ can be obtained, and hence, $\hat{\boldsymbol{P}}_{spot,q_i}$ can be resolved.

\vspace{-2mm}
\section{Proof of Lemma \ref{Lemma4}}\label{Appendix4}
Let $d_{i^*}= \sqrt{(x_{i^*}-x_{len})^2+(y_{i^*}-y_{len})^2+(z_{i^*}-z_{len})^2}$ denotes the distance to the closest LED from the lens, where $(x_{i^*}, y_{i^*},z_{i^*})$ as the coordinate of the selected $i^*$-th LED. Equating coefficients along $x$, $y$, and $z$ directions in $\hat{\boldsymbol{\eta}}_{i^*,l}$ to $\hat{\boldsymbol{\eta}}_{len}$ the following equations hold,
\vspace{-1mm}
\begin{align}\label{eqs:36}
\hspace{-3.4mm}
\resizebox{0.44\textwidth}{!}{$
\text{c}{\theta_R}\text{c}{\phi_R}\text{c}{\theta_L}\text{s}{\phi_L}-\text{s}{\theta_R}\text{s}{\theta_L}\text{s}{\phi_L}+
\text{c}{\theta_R}\text{s}{\phi_R}\text{c}{\phi_L} = \frac{x_{i^*}-x_{len}}{d_{i^*}}$}, 
\end{align}
    \vspace{-7mm}
\begin{align}\label{eqs:37}
\hspace{-3.4mm}
\resizebox{0.44\textwidth}{!}{$
\text{s}{\theta_R}\text{c}{\phi_R}\text{c}{\theta_L}\text{s}{\phi_L}+\text{c}{\theta_R}\text{s}{\theta_L}\text{s}{\phi_L}+
\text{s}{\theta_R}\text{s}{\phi_R}\text{c}{\phi_L} = \frac{y_{i^*}-y_{len}}{d_{i^*}}$},
\end{align}
    \vspace{-7mm}
\begin{align}\label{eqs:38}
\hspace{-3.4mm}
\resizebox{0.27\textwidth}{!}{$
-\text{s}{\phi_R}\text{c}{\theta_L}\text{s}{\phi_L}+\text{c}{\phi_R}\text{c}{\phi_L} = \frac{z_{i^*}-z_{len}}{d_{i^*}}$}.
\end{align}
The expressions in \eqref{eqs:36}, \eqref{eqs:37}, and \eqref{eqs:38} can be combined to obtain closed-form expressions for $\theta_L$ and $\phi_L$ for this scheme. For this purpose, \eqref{eqs:36} is multiplied with $\text{s} \theta_R$ and \eqref{eqs:37} is multiplied with $\text{c} \theta_R$. The resulting expressions can be subtracted to get a single expression. The resulting expression and \eqref{eqs:38} are combined, and after mathematical manipulations, a quadratic equation of the variable $\text{c}{\phi_L}$ can be expressed as
\vspace{-3mm}
    \begin{align}\label{eqs:quad1}
    \hspace{-3.4mm}
    &\resizebox{0.44\textwidth}{!}{$\text{c}{\phi_L}^2-2\text{c}{\phi_R}\left(\frac{z_{i^*}-z_{len}}{d_{i^*}}\right)\text{c}{\phi_L} + \left(\frac{z_{i^*}-z_{len}}{d_{i^*}}\right)^2 
    + \text{s}{\phi_R}^2\times$}\nonumber \\
    &\resizebox{0.35\textwidth}{!}{$\left(\text{c}{\theta_R} \hspace{-0.5ex} \left(\frac{y_{i^*}-y_{len}}{d_{i^*}}\right)\hspace{-0.5ex}-\hspace{-0.5ex} \text{s}{\theta_R}\left(\frac{x_{i^*}-x_{len}}{d_{i^*}}\right)\right)^2
    \hspace{-0.5ex}- \hspace{-0.5ex}\text{s}{\phi_R}^2\hspace{-0.5ex}= \hspace{-0.5ex}0$}.
    \vspace{-3mm}
    \end{align}
After finding the solutions for the quadratic equation in \eqref{eqs:quad1}, $\text{c}{\phi_L}$ is expressed as
\vspace{-3mm}
    \begin{align}\label{eqs:quad1_sol}
    &\resizebox{0.32\textwidth}{!}{$\text{c}{\phi_L} = \text{c}{\theta_R}\left(\frac{z_{i^*}-z_{len}}{d_{i^*}}\right)\pm\text{s}{\theta_R}\Big(1-\left(\frac{z_{i^*}-z_{len}}{d_{i^*}}\right)^2$}\nonumber \\
    &\resizebox{0.32\textwidth}{!}{$-\left(\text{c}{\theta_R} \left(\frac{y_{i^*}-y_{len}}{d_{i^*}}\right)- \text{s}{\theta_R}\left(\frac{x_{i^*}-x_{len}}{d_{i^*}}\right)\right)^2\Big)^{\frac{1}{2}}$}.
        \vspace{-7mm}
    \end{align}
However, from the geometrical observations it is noted that the square-root component of \eqref{eqs:quad1_sol} is zero and $\phi_L$ has a distinct solution, which is the final expression for $\phi_L$ in Lemma \ref{Lemma3}. Finally, by using \eqref{eqs:36}, \eqref{eqs:37}, and \eqref{eqs:quad1_sol_sim}, with the relation $\sin{\theta} = \sqrt{1-\cos^2{\theta}}$, the final expression for $\theta_L$ can be derived.

\vspace{-2mm}
\bibliographystyle{IEEEtran}
\bibliography{IEEEabrv,Main}

\begin{thebibliography}{10}
\providecommand{\url}[1]{#1}
\csname url@samestyle\endcsname
\providecommand{\newblock}{\relax}
\providecommand{\bibinfo}[2]{#2}
\providecommand{\BIBentrySTDinterwordspacing}{\spaceskip=0pt\relax}
\providecommand{\BIBentryALTinterwordstretchfactor}{4}
\providecommand{\BIBentryALTinterwordspacing}{\spaceskip=\fontdimen2\font plus
\BIBentryALTinterwordstretchfactor\fontdimen3\font minus \fontdimen4\font\relax}
\providecommand{\BIBforeignlanguage}[2]{{%
\expandafter\ifx\csname l@#1\endcsname\relax
\typeout{** WARNING: IEEEtran.bst: No hyphenation pattern has been}%
\typeout{** loaded for the language `#1'. Using the pattern for}%
\typeout{** the default language instead.}%
\else
\language=\csname l@#1\endcsname
\fi
#2}}
\providecommand{\BIBdecl}{\relax}
\BIBdecl

\bibitem{Ghassemlooy}
Z.~Ghassemlooy, S.~Arnon, M.~Uysal, Z.~Xu, and J.~Cheng, ``Emerging optical wireless communications-{Advances} and challenges,'' \emph{IEEE J. Sel. Areas Commun.}, vol.~33, pp. 1738--1749, Sep. 2015.

\bibitem{Fath_2023}
T.~Fath and H.~Haas, ``Performance comparison of {MIMO} techniques for optical wireless communications in indoor environments,'' \emph{{IEEE} Trans. Commun.}, vol.~61, pp. 733--742, Feb. 2013.

\bibitem{chockalingam_2015}
S.~P. Alaka, T.~L. Narasimhan, and A.~Chockalingam, ``Generalized spatial modulation in indoor wireless visible light communication,'' in \emph{Proc. IEEE GLOBECOM 2015}, San Diego, CA, Dec. 2015, pp. 1--7.

\bibitem{Nahhal_2021}
M.~Al-Nahhal, E.~Basar, and M.~Uysal, ``Flexible generalized spatial modulation for visible light communications,'' \emph{{IEEE} Trans. Veh. Technol.}, vol.~70, pp. 1041--1045, Jan. 2021.

\bibitem{Wang_2018}
T.~Wang, F.~Yang, L.~Cheng, and J.~Song, ``Spectral-efficient generalized spatial modulation based hybrid dimming scheme with {LACO-OFDM} in {VLC},'' \emph{{IEEE} Access}, vol.~6, pp. 41\,153--41\,162, Jun. 2018.

\bibitem{Chen_2021}
Y.~Chen, S.~Gao, G.~Tu, and D.~Chen, ``Group-based {LED} selection for generalized spatial modulation in visible light communication,'' \emph{{IEEE} Commun. Lett.}, vol.~25, pp. 3022--3026, Sep. 2021.

\bibitem{Asanka_2015}
A.~Nuwanpriya, S.-W. Ho, and C.~S. Chen, ``Indoor {MIMO} visible light communications: Novel angle diversity receivers for mobile users,'' \emph{{IEEE} J. Sel. Areas Commun.}, vol.~33, pp. 1780--1792, Sep. 2015.

\bibitem{Shiyuan_2023}
S.~Sun, F.~Yang, J.~Song, and R.~Zhang, ``Intelligent reflecting surface for {MIMO} {VLC}: Joint design of surface configuration and transceiver signal processing,'' \emph{{IEEE} Trans. Wirel. Commun.}, vol.~22, pp. 5785--5799, Sep. 2023.

\bibitem{Sushanth_2018}
K.~V. S.~S. Sushanth and A.~Chockalingam, ``Performance of imaging receivers using convex lens in indoor {MIMO} {VLC} systems,'' in \emph{Proc. {IEEE} VTC-Spring 2018}, Porto, Portugal, Jun. 2018, pp. 1--5.

\bibitem{Ngatched_2021}
A.~R. Ndjiongue, T.~M.~N. Ngatched, O.~A. Dobre, and H.~Haas, ``Re-configurable intelligent surface-based {VLC} receivers using tunable liquid-crystals: The concept,'' \emph{J. Light. Technol.}, vol.~39, pp. 3193--3200, May 2021.

\bibitem{Zohrabi_2016}
M.~Zohrabi, R.~H. Cormack, and J.~T. Gopinath, ``Wide-angle nonmechanical beam steering using liquid lenses,'' \emph{Opt. Express}, vol.~24, pp. 23\,798--23\,809, Oct. 2016.

\bibitem{Tian_2022}
J.-Q. Tian, Z.-Z. Zhao, and L.~Li, ``Adaptive liquid lens with a tunable field of view,'' \emph{Opt. Express}, vol.~30, pp. 40\,991--41\,001, Oct. 2022.

\bibitem{Bart_1986}
B.~Braden, ``The surveyor's area formula,'' \emph{College Math. J.}, vol.~17, pp. 326--337, 1986.

\bibitem{Matthiesen_2020}
B.~Matthiesen, C.~Hellings, E.~A. Jorswieck, and W.~Utschick, ``Mixed monotonic programming for fast global optimization,'' \emph{{IEEE} Trans. Signal Process.}, vol.~68, pp. 2529--2544, Mar. 2020.

\bibitem{Amit_2012}
A.~Agrawal, S.~Ramalingam, Y.~Taguchi, and V.~Chari, ``A theory of multi-layer flat refractive geometry,'' in \emph{Proc. {IEEE} CVPR 2012}, Providence, USA, 2012, pp. 3346--3353.

\end{thebibliography}

\end{document}